# Load Frequency Control For Distributed Grid Power System Single Area & Multi-area System

Sushant Chaudhary
M-Tech Scholar, Control System Engg. , Amity University, Noida, India.
sushant0561@gmail.com

*Abstract*— **This project presents decentralized control scheme for Load-Frequency Control in Multi-area power System. In this era renewable energy is most promising solution to man's ever increasing energy needs. But the power production by these resources cannot be controlled unlike in thermal plants. A number of optimal control techniques are adopted to implement a reliable stabilising controller. It is necessary to interconnect more distributed generation in power systems because of environmental concerns. Primary concern includes global environmental and energy depletion problem. A serious attempt has been undertaken aiming at investigating the load frequency control problem in a power system consisting of two power generation unit and multiple variable load units. The robustness and reliability of the various control schemes is examined through simulations. A system involving thermal plants and a hydro plant is modelled using MATLAB 2011b.**

## I. INTRODUCTION

Load frequency control (LFC) in power system maintains frequency deviation & tie-line power flow deviation, which constantly vary according to load and disturbance with in preset values. In India majority of power production, around 65% is from thermal plants. Due to problem related to uncertainty, pricing and supply of fossil fuels renewable resources have been identified as best alternative [1]. The fitful nature of resources increases the frequency deviation which further adds to deviation caused by Load variation. This necessitates the grid connection of renewable resources [2]. For large scale power systems which consists of inter-connected control areas, load frequency then it is important to keep the frequency and inter area tie power near to the scheduled values. However one concern is that large, short term output fluctuations may influence system frequency because it is difficult to estimate renewable energy output. Frequency deviation is undesirable because most of AC motors are directly related to frequency. Also the generator turbines are designed to operate at precise speed. This is imperative to maintain constant frequency. This is done by implementing Load Frequency Control (LFC). There are many LFC methods developed for controlling frequency [3]. The load frequency control mainly deals with the control of the system frequency and real power whereas the automatic Voltage regulator loop regulates the changes in the reactive power and voltage magnitude. Load frequency control is the basis of many advanced concepts of the large scale control of the power system.

## II. MODELLING

1. Modelling of Generator :-

Applying the swing equation of synchronous machine to small perturbation we have,

$$\frac{2H}{\omega}\frac{d^2\delta}{dt^2} = \Delta P_m - \Delta P_e \qquad (1)$$

Taking Laplace transform we get,

$$\Delta\Omega(s) = \frac{1}{2Hs}[\Delta P_m(s) - \Delta P_e(s)] \qquad (2)$$

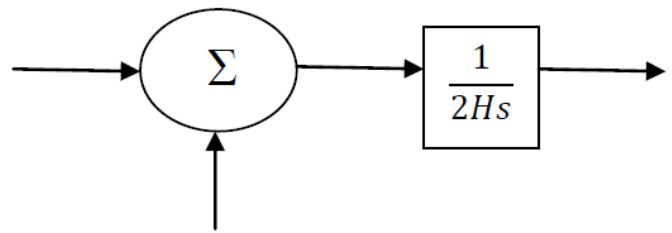

Fig 1. Block representation of Generator

2. Mathematical modelling of Load :-

The load on power system consists of variety of electrical drives. The speed-Load characteristic of composite load is given by:-

$$\Delta P_e = \Delta P_L - D\Delta\omega \qquad (3)$$

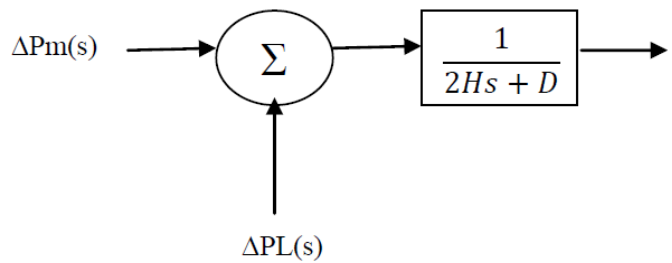

Fig 2. Fig 1. Block representation of Load

3. Mathematical modelling of prime mover :-

The sources of power generation are commonly known as the prime mover.

$$\Delta G_T = \frac{\Delta P_m(s)}{\Delta P_v(s)} = \frac{1}{\tau(s)+1} \qquad (4)$$

4. Mathematical Modelling of governor:-

When the electrical load is suddenly increased then the electrical power exceeds the mechanical power input. As a result of this the deficiency of power in the load side is extracted from the rotating energy of the turbine. Due to this reason the kinetic energy of the turbine i.e. the energy stored in the machine is reduced and the governor sends a signal to supply more volumes of water or steam or gas to increase the speed of the prime-mover so as to compensate speed deficiency.

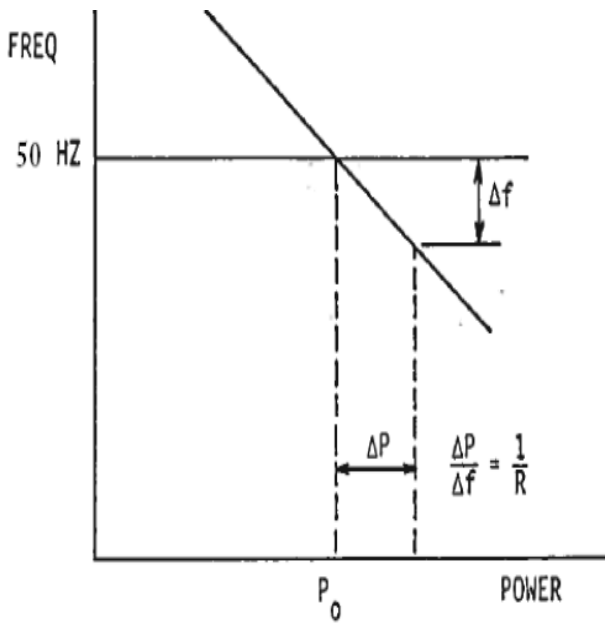

Fig 3. Frequency Deviation

The slope of the curve represents speed regulation R. Governors typically have a speed regulation of 5-6 % from no load to full load

5. Entire Thermal Area :-

The thermal model have been modelled by using transfer function[5].

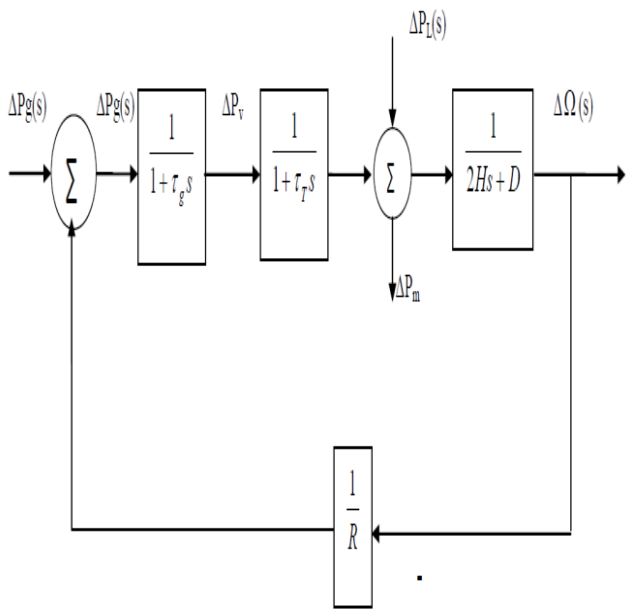

Fig 4. Mathematical modelling of block diagram of single area system.

6. Modelling of Hydro Power Plant

The representation of the hydraulic turbine and water column in stability studies is usually based on certain assumptions. The hydraulic resistance is considered negligible. The penstock pipe is assumed inelastic and water incompressible. Also the velocity of the water is considered to vary directly with the gate opening and with the square root of the net head and the turbine power output is directly proportional to product of head and volume flow(4). Hydro project are similarly modelled as thermal plants[7].

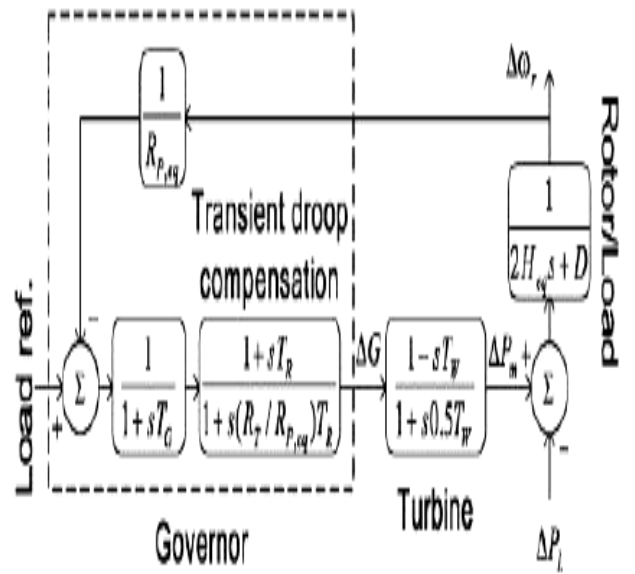

Fig 5. Block diagram of hydro area

### III. LFC FOR SINGLE AREA:-

With the primary LFC loop a change in the system load will result in a steady state frequency deviation , depending on the governor speed regulation. In order to reduce the frequency deviation to zero we must provide a reset action by introducing an integral controller to act on the load reference setting to change the speed set point. The integral controller increases the system type by 1 which force the final frequency deviation to zero. The integral controller gain must be adjusted for a satisfactory transient response[6].

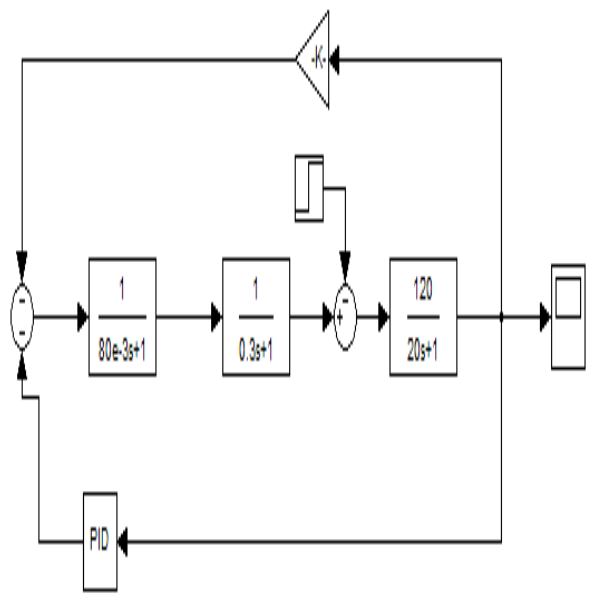

Fig 6. Proportional plus integral Load frequency control.

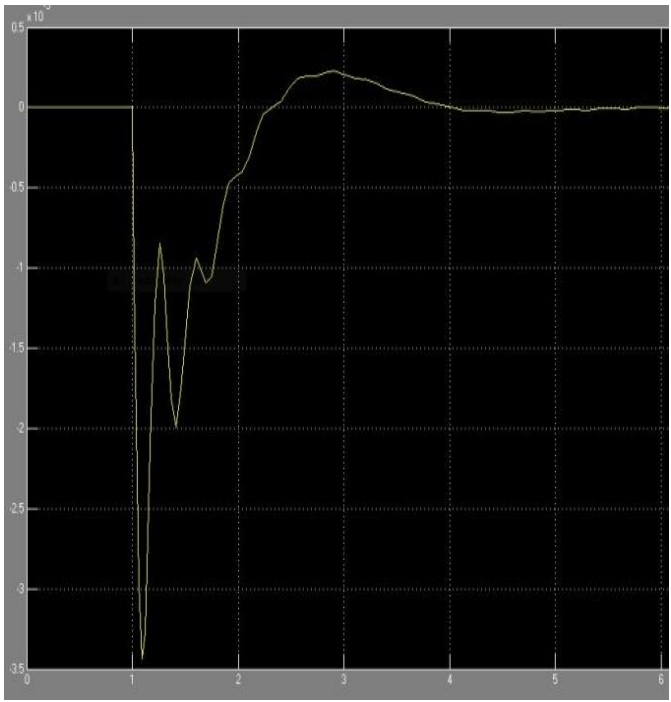

Fig:7.Simulink Graph for single area

## IV: SIMULATION AND RESULTS

1. LFC for thermal area

A thermal system with pid load frequency control is simulated in Fig 7.

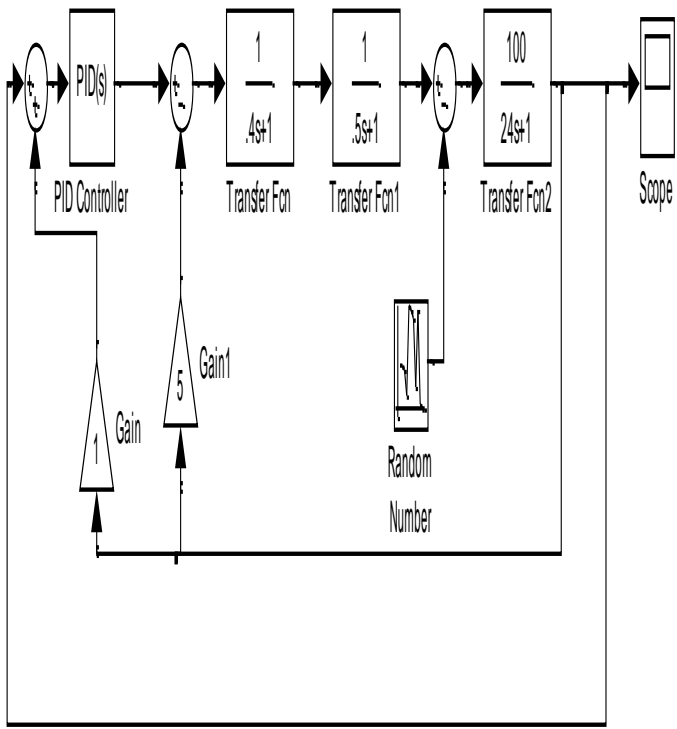

Fig 8. Simulink for thermal system.

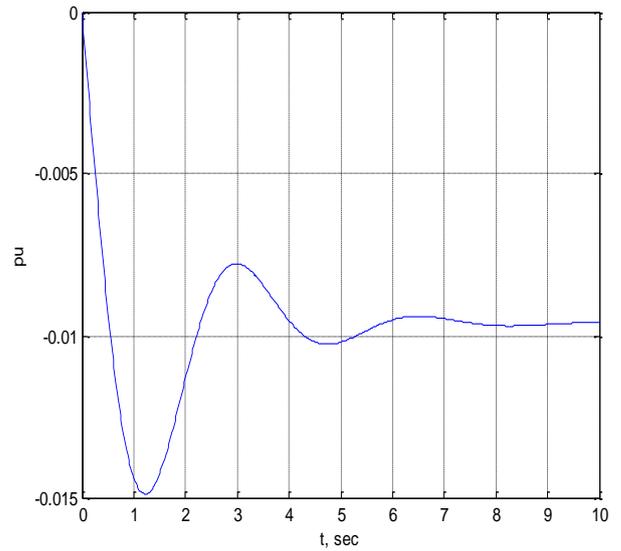

Fig 9. Response of fixed load.

2. LFC for Hydro area

Multi-area load frequency model for Hydro power plant is simulated in fig 9.

Frequency deviation versus time for integrated for hydro system with step load change is shown in fig 10[8].

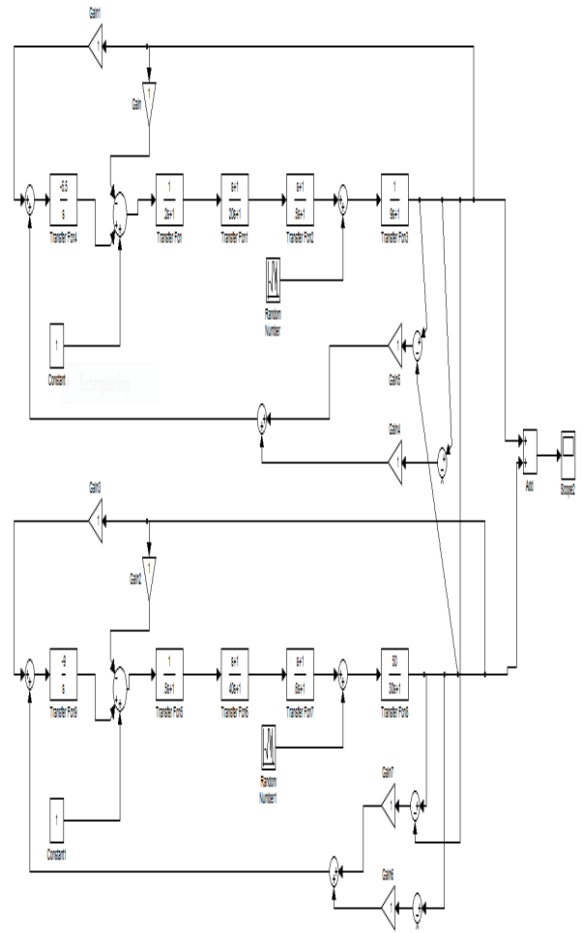

Fig 10. Simulink for multi-area hydro system

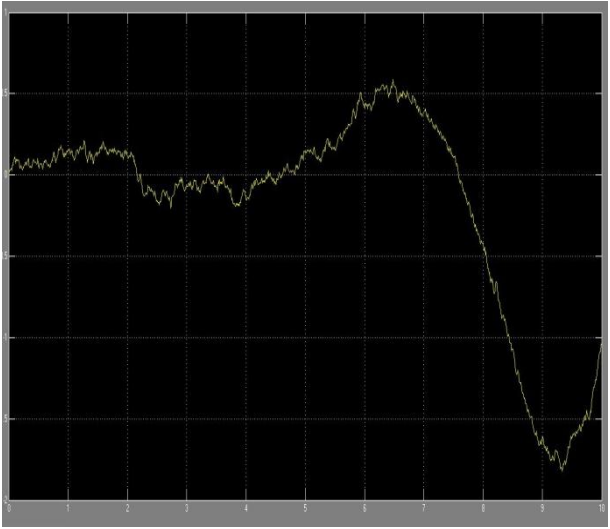

Fig11. Step response for uncompensated system

From the above fig it is clearly noticed that hydro energy does not effect the system frequency frequency deviation is just within limits.

3.LFC for Thermal and Hydro System

The two hydro system along with thermal system is combined and simulated diagram is shown in fig 11. From fig 12 it is clearly shown that thermal system does not effect the frequency deviation.

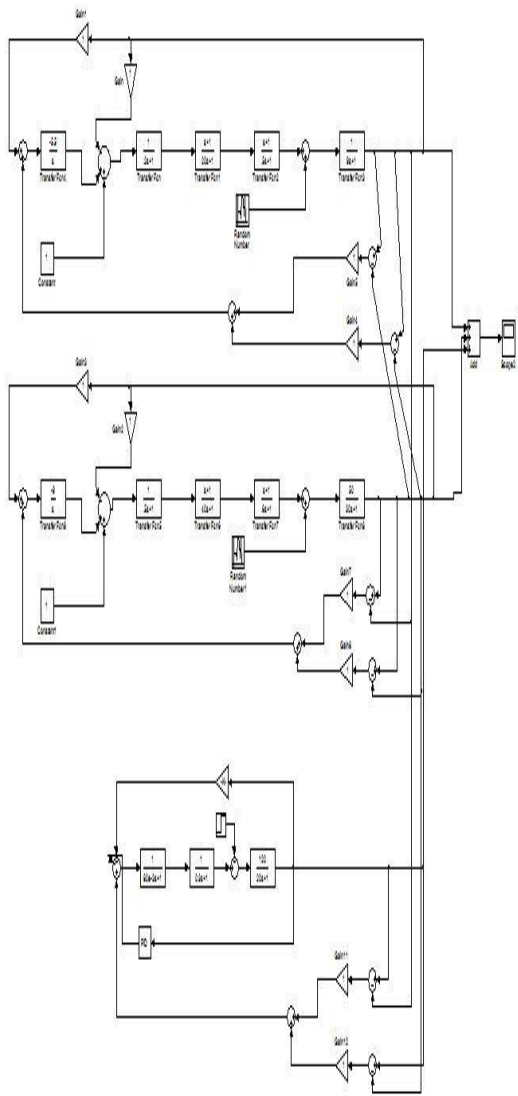

Fig12. Thermal and Hydro system

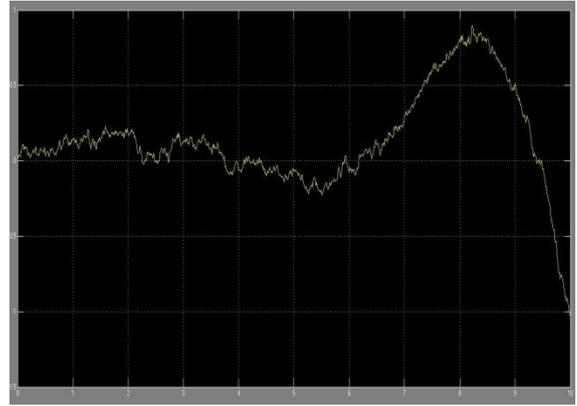

Fig 13. Simulink graph for combined system

## V. CONCLUSION

The project presents a case study of designing a controller that can bear desirable results in a two area power system when the input parameters to the system is changed. Load frequency control becomes more important when a large amount of renewable energy sources are introduced. In this paper load frequency control of with considerable penetration of renewable has been analysed and frequency deviation is found within tolerable limits.